\documentclass{article}
\usepackage{spconf,amsmath,graphicx}

\usepackage{color, bm}
\usepackage{url}

\usepackage[hang,flushmargin]{footmisc}

\renewcommand{\footnoterule}{%
  \kern -3pt
  \hrule width \columnwidth height 0.4pt 
  \kern 2.6pt
}

\title{PUFFIN: PITCH-SYNCHRONOUS NEURAL WAVEFORM GENERATION FOR FULLBAND SPEECH ON MODEST DEVICES}

\name{Oliver Watts$^*$, Lovisa Wihlborg$^*$, Cassia Valentini-Botinhao$^*\dagger$}

\address{
$^*$SpeakUnique Ltd.\\
$^\dagger$CSTR, University of Edinburgh, UK 
}

\begin{document}
\ninept

\maketitle

\begin{abstract} 
We present a neural vocoder designed with low-powered Alternative and
Augmentative Communication  devices in mind. By combining elements of
successful modern vocoders with established ideas from an older generation of
technology, our system is able to produce high quality synthetic speech at
48kHz on devices where neural vocoders are otherwise prohibitively complex. The
system is trained adversarially using differentiable pitch synchronous overlap
add, and reduces complexity by relying on pitch synchronous Inverse Short-Time
Fourier Transform (ISTFT) to generate speech samples. Our system achieves
comparable quality with a strong baseline (HiFi-GAN) while using only a
fraction of the compute. We present results of a perceptual evaluation as well
as an analysis of system complexity.
\end{abstract}


\section{Introduction}

The rapid improvements in quality of text-to-speech (TTS) synthesis in recent
years have been due in large part to new methods of waveform generation based on
artificial neural networks. In contrast to older approaches \cite{Masanori16, 
Espic17} where acoustic features are transformed into a speech
waveform via fixed, knowledge-based procedures, recent work has successfully
replaced these fixed transformations with ones learned from data \cite
{mehri2016samplernn, shen2017natural}. This recent progress has resulted in
several designs of trainable vocoder generating near-human quality speech. They
have seen wide adoption, the training of these is tractable and the models can
synthesise in real time given powerful CPUs \cite{matsubara2022comparison}.

A useful application of TTS is in Alternative and Augmentative Communication
(AAC), used by people with speech and/or language impairments. Ideally, AAC
users should be able to benefit from the increase in quality that neural
waveform generation has achieved lately; however, often AAC devices are not
powerful enough for this to be possible. Many such devices are based around
relatively low-powered CPUs with clock speeds that are low and SIMD registers
that are small compared with the hardware in many recent smartphones. While
these CPUs have advantages such as low power consumption, they present serious
challenges for neural vocoders: even state-of-the-art models that have been
designed specifically for efficiency are not viable on such hardware. 

Operation of the first successful neural waveform generators \cite
{oord2016wavenet} was slow due to the inherently high frequency of audio data,
and also to the fact that they operated autoregressively, conditioning the
prediction of each sample on previous samples generated. One approach to
speeding up audio generation has been to develop models which generate waveform
samples in parallel without the expense of making predictions
autoregressively \cite{oord2018parallel, prenger2019waveglow}. One line of work
has looked at training such parallel models adversarially \cite
{kumar2019melgan, hifigan}. Other ways of improving the speed of neural
waveform generation include incorporating ideas from signal processing (such as
linear prediction) into machine learning models \cite{valin2019lpcnet}, and
multiband modelling to reduce the rate at which speech must be generated by
generating several bands in parallel \cite{cui2020efficient}. Many systems take
advantage of the possibility of having parts of a model which operate at a rate
much lower than the speech sampling rate, see e.g.\ the frame-level
conditioning network in LPCNet \cite{valin2019lpcnet} or the gradual upsampling
used in HiFi-GAN \cite{hifigan}. Those models still have elements that must
operate at the output sampling rate, however; \cite{kaneko2022istftnet} and \cite{webber2023autovocoder} go
further in replacing a number of HiFi-GAN's faster-rate layers with what is
essentially a deterministic upsampling operation, the Inverse Short-Time
Fourier Transform (ISTFT). 

We present a neural vocoder designed with low-powered AAC devices in mind. The
approach presented here builds on several of the trends mentioned above.
Combining elements of successful modern vocoders (such as adversarial training)
with established ideas from an older generation of vocoders (such as pitch
synchronous processing) allows high quality synthetic speech to be generated on
low-powered AAC devices. We achieve comparable quality with a strong
baseline (HiFi-GAN) \cite{hifigan}, using only a fraction of the compute
to generate speech at over twice the sampling rate (48kHz).\footnote{
{\footnotesize Samples: \ttfamily \url
{https://speakunique.github.io/puffin_demo/}}}

\section{Proposed system}\label{proposed_system}

\begin{figure}[t]
\vspace{-8pt}
\includegraphics[scale=0.55]{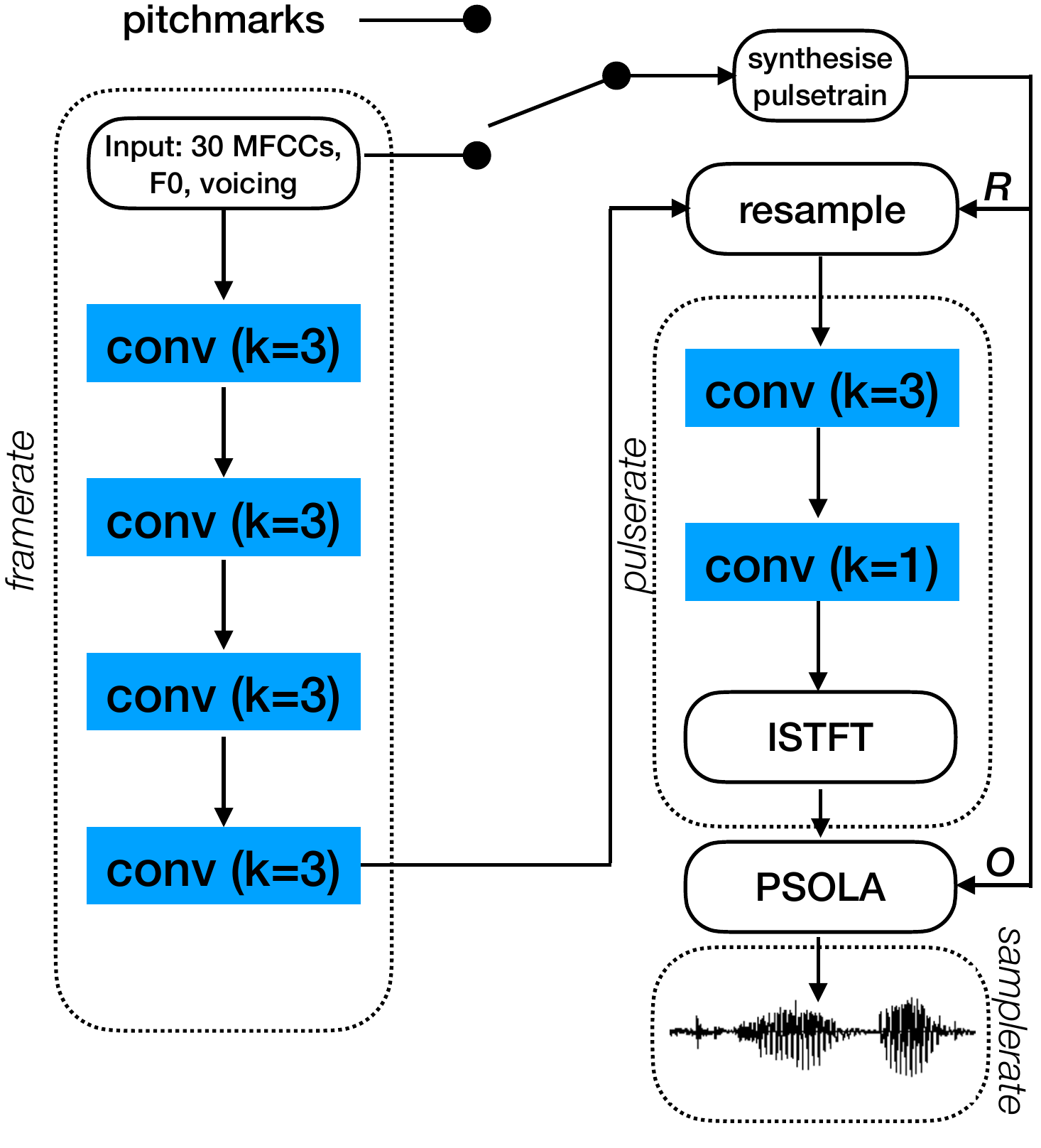}
\vspace{-8pt}
\caption{Overview of the Puffin generator.} \label{fig:sysdiag}
\vspace{-6pt}
\end{figure}

We propose a system whose submodules operate at 3 different rates, as illustrated by Figure \ref{fig:sysdiag}: a fixed input
frame rate (100Hz), a variable pulse rate which depends on the $F_0$ and
voicing of speech (average 131Hz, maximum 400Hz), and output audio sample rate
(48000Hz). Importantly, the system is able to efficiently generate such
wideband audio -- covering all frequencies which can be perceived by human
listeners -- due to the fact that neural network operations are performed only
at the first two much slower rates. As in \cite{kaneko2022istftnet} an output
at the desired sample rate is obtained using ISTFT, but here this is done pitch
synchronously, and with successful use of much greater FFT window lengths and
shifts, drastically reducing the computation required to generate speech at
more than twice the sample rate. 

\begin{figure}[t]
\vspace{-8pt}
\includegraphics[scale=0.45]{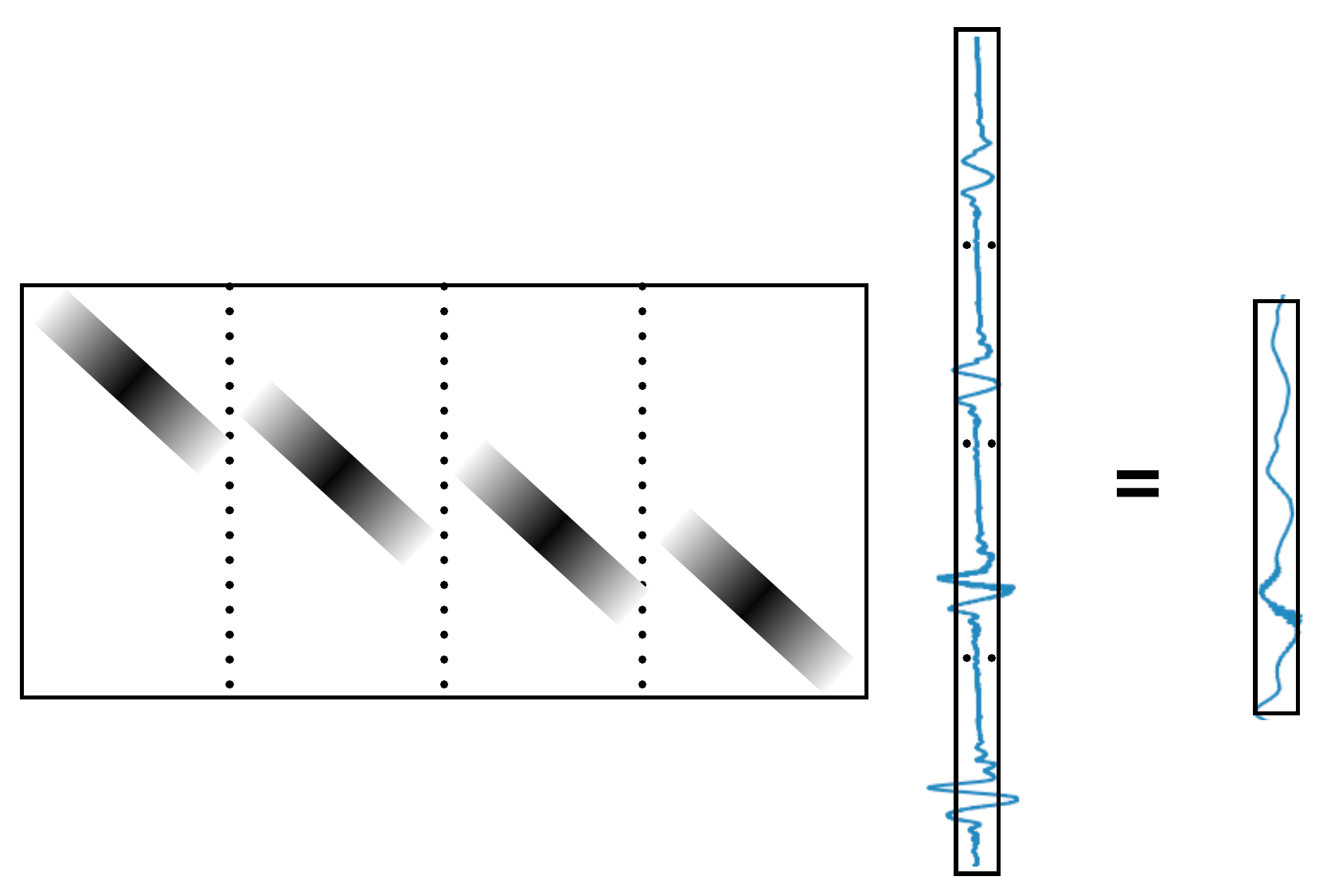}
\vspace{-8pt}
\caption{Overlap add. On the left the $\bm{O}$ matrix.} \label{fig:ola}
\vspace{-8pt}
\end{figure}

The following notation is used here: $T$ is number of fixed rate timesteps to be
processed by the model; $P$ is the number of glottal pulses in the same
example; $F$ is an FFT length which we set to be wider than the largest pulses
observed (2048). $T$ is fixed to 512 per example in training and the
corresponding $P$ depends on the $F_0$ of the speech in the example.

The frame-rate part of the
network passes 32 dimensional inputs at 10 ms intervals through 4 simple 1D
convolutional layers, each  with kernel width $k$ equal to 3, and each followed by
leaky ReLU activation. The 32 input features consist of 30 mel frequency
cepstral coefficients together with $F_0$ (linearly interpolated through
unvoiced regions) and voicing. The features are all scaled to appropriate
magnitudes using global constants. This results in an array of hidden
activations of size T$\times$H. (H=256 for standard setting, and H=1024 for
large as explained below.) 

As shown in Figure \ref{fig:sysdiag}, $F_0$-related data is used to map between
the 3 different rates, using matrix multiplications. Based on an $F_0$ track at
synthesis time --  $F_0$ is included among the input features -- and on a
pitchmark track in training (which specifies the locations of detected glottal
closures in the training audio), a resampling matrix $\bm{R}$ and an
overlap-add matrix $\bm{O}$ are generated. 

Multiplication of fixed rate data with $\bm{R}$ resamples it to the pulse rate
by linear interpolation, such that time steps are no longer aligned with fixed
rate analysis windows used to provide the input, but instead centered on
glottal closure locations. The operation is broadcast across channels, such
that channels are interpolated independently of one another. The resampled
hidden activations are then  processed by a further width-3 convolutional layer
followed by leaky ReLU. The final layer with learned parameters is a
time-distributed feedforward layer (i.e.\ a convolutional layer with kernel
width 1); its job is to increase the number of channels so that the data can be
split into two $F+1$ portions
which can be treated as the real and imaginary parts of a pitch-synchronised
complex spectrogram. An inverse Fast Fourier Transform (IFFT) is used to
convert each slice of this spectrogram into an $F$-dimensional fragment of
speech waveform.  
Following \cite[\S2.3.3]{Espic17}, each of these fragments is treated as a circular buffer whose samples are shifted forward by $F/2$. Although in our system there is no explicit delay compensation step that must be reversed as in \cite[\S2.2.3]{Espic17}, motivation is similar as it is expected that backpropagating through this correction operation will result in intermediate representations of phase that evolve smoothly over time.
The fragments are concatenated end-to-end in a PF x 1
dimensional array. $\bm{O}$ is a TS$\times$PF matrix (where S=480, the number of
samples in a fixed-rate frame), which is constructed such that the matrix
product between it and the concatenated data yields the waveform fragments
assembled by overlap-add into a speech waveform of the correct duration.
Figure \ref{fig:ola} illustrates this operation schematically. The windows are
centered on the positions of glottal closures in the output waveform. We use assymetric Hann windows that extend to the two
neighbouring glottal closure positions, similar to \cite{Espic17}. There is no strict need to use tapering
windows of this kind -- in principle rectangular windows are adequate as the
waveform fragments that are being assembled are an output of a neural network
which can learn to provide data with the appropriate characteristics. However,
tests early on in development suggested that using a tapered window is
beneficial here.  

Note that while implementing resampling and overlap add with matrix
multiplication requires the use of matrices covering a sentence or a batch,
this is used for the purposes of optimisation only. For deployment these
operations are implemented in a way that allows incremental streaming synthesis
with minimal lookahead. 

$\bm{R}$ and especially $\bm{O}$ are both very large and sparse matrices, and training is
only possible using sparse matrix operations. Standard sparse serialisation
formats are not adequate to store $\bm{O}$ efficiently but its structure can be exploited to devise an efficient custom format.

\subsection{Network training} As with other adversarially trained waveform
 generators (e.g.\ \cite{hifigan}), we train the generator using a combination
 of GAN losses and $L_1$ losses. We use a least squares GAN approach as
 in \cite{hifigan} which is similar in terms of training setup except from the
 architectures of the discriminators themselves -- our discriminators all
 operate in the frequency domain. They take the time domain signal output by
 the generator shown in Figure \ref{fig:sysdiag}, and convert it to a complex
 spectrogram, whose two channels contain real and imaginary components. Each
 subdiscriminator contains a stack of five 2D convolutional layers,
 interspersed with leaky ReLUs and each having a kernel of width 3 along both
 time and frequency axes (except for the last layer having a kernel size of 1 $\times$ 3). The subdiscriminators therefore have a receptive
 field covering 9 timesteps and 11 frequency bins. We found it advantageous to
 use an ensemble of subdiscriminators, covering a number of different analysis
 window lengths and shifts (in the range 128--4096 points and 256--1024 points,
 respectively). Furthermore, we saw gains from having such an ensemble
 specialising in different spectral bands, and assigned 3, 3, and 2 submodels
 to each of the 3 8kHz frequency bands that make up the 24kHz bandwidth captured by
 the 48kHz sampling rate. 

We also make use of an ensemble of $L_1$ losses between natural and generated
magnitude spectrograms. As well as a mel spectral loss as used in  \cite
{hifigan} we found it useful to also use $L_1$ losses computed from linear
magnitude spectrograms extracted with a variety of 6 different analysis window
shifts/length configurations \cite{wangNSF, yamamoto2020parallel}.

The contribution of each $L_1$ loss was weighted by 0.5 and combined with the
adversarial loss terms. In contrast to \cite{hifigan}, no feature
matching loss was used.

It was found effective to train for a number of steps using only $L_1$ losses,
which can be done quickly and results in a generator which produces speech that
is clear but marred by phase artefacts. Further training using the GAN losses
for generator and discriminators reduces these artefacts. 

\vspace{-10pt}
\section{Experiments}
\label{sec:experiments}

\begin{table}[t]
\centering
\setlength{\tabcolsep}{3pt} 
\caption{Experimental conditions.\label{system_table}}
\begin{tabular}{ l  l   }  \hline
System  & Description  \\ \hline
N  & Natural speech  \\ 
H$^1$ & HiFi-GAN v1 \cite{hifigan}  \\
H$^3$ & HiFi-GAN v3 \cite{hifigan}  \\
P & Proposed  \\ 
P$^D$ & Proposed (\textbf{D}ownsampled to 22.05kHz) \\ 
P$^L$ & Proposed (\textbf{L}arger)  \\
\hline
\end{tabular}
\vspace{-18pt}
\end{table}

\vspace{-10pt}
\subsection{Database}
\vspace{-5pt}
In order to compare our system with a publicly available HiFi-GAN model we use
the VCTK dataset \cite{vctk}. This dataset contains around 20mins of data from
109 native English speakers. The data is available at 96kHz which allows
training of 48kHz models. 

From the VCTK dataset we held out 9 speakers from training, following the setup
of \cite{hifigan} as closely as possible. Exact details of the train/test split
used by \cite{hifigan} are not available, but from the demo samples associated
with that work\footnote{\url{https://jik876.github.io/hifi-gan-demo/}} we infer
the identity of 4 of the  nine held-out speakers (p226, p271, p292,
p318) excluded from the HiFi-GAN model
training. We excluded 5 other speakers (p225, p234, p236, p238,
p239) from our training data so that its
quantity is comparable to that used to train the existing HiFi-GAN models. To
supplement the 4 VCTK evaluation speakers found in this way, we added data from
2 other publicly available single-speaker databases recorded in the same
conditions as VCTK  \cite{NickData, AlbaData}.

All train and evaluation sentences were endpointed, normalised for
loudness \cite{sv56} and pitchtracked using \cite{Reaper} (minimum and maximum
values for $F_0$ were set to 50Hz and 400Hz).

\vspace{-8pt}
\subsection{Proposed system}

\vspace{-5pt}
The system described in Section 2 was trained using  2 configurations. The
standard configuration P sets number of channels in hidden activations $H$ to
256 and applies block sparsity to the large final layer \cite
{valin2019lpcnet}, retaining only 10\% of the weights. We also trained a large
system with more parameters P$^{L}$ ($H$=1024) and no sparsity in order to be
able to evaluate a Puffin system with comparable complexity to baseline system
H$^3$ (see Table \ref{system_table}). Furthermore, Puffin operates at a higher sample rate than the baseline,
and to evaluate the contribution of the upper frequency band to perceived
quality we downsampled system P's samples to 22,050Hz and included them as
condition P$^D$.

Training was done with batches of 512 frames of speech concatenated in the time
axis. The Adam optimiser was used. We trained both P and P$^L$ for 300,000
steps (c.3.75 hours on a single GTX1080Ti for P) using $L_1$ loss terms only; over this interval, desired sparsity was
introduced into P using the schedule described in \cite{valin2019lpcnet}. A
further 100,000 steps of training (c.26 hours on the same hardware for P) were carried out for both models, including
GAN loss terms, and maintaining the sparsity of system P.

\vspace{-5pt}
\subsection{Baseline systems}
\vspace{-5pt}

We included two versions of HiFi-GAN \cite{hifigan} that differ in terms of the
generator's architecture: H$^1$ is version 1 (tuned for quality) and H$^1$
is version 3 (tuned for quality/performance trade-off). Published 
code and models were used to produce stimuli for these systems.\footnote{\url
{https://github.com/jik876/hifi-gan}}

\vspace{-10pt}
\subsection{Complexity}
\vspace{-5pt}

We compute approximate complexity of each evaluated model by counting the number
of operations performed in a forward pass to generate 1 second of speech. This
is dominated by the multiplication and addition that must be performed for each
weight in the network at the relevant rate. Following \cite{valin2019lpcnet} we
omit computation associated with biases, activation functions, ISTFT etc. --
empirically we find that the latter contributes only slightly to overall run
time. Calculation of the complexity contribution of each module of each of the 4
systems is given in Table \ref{eff_table2}.  

All modules of all 4 systems are made up of convolutional layers
(some transposed, and some used to compute residuals). The essential complexity
of a convolutional layer is taken as $iok$ ($i$: number of input channels, $o$:
output channels, $k$: kernel width) and is not affected by stride and dilation
(see columns 2--4 of Table \ref{eff_table2}). Residual blocks multiply this by
number of convolutional layers used $l$; where sparsity is used, complexity is
multiplied by  fraction of retained weights $d$ (column 5 of Table \ref
{eff_table2}). This is then multiplied by frequency of operation (column 6) and
finally multiplied by 2 (to account for both multiply and add operations).
Therefore column 7 of Table \ref{eff_table2} shows twice the product of
previous columns. Approximate total MFLOPS per system are given after each section.

We note that Puffin's efficiency depends on the $F_0$ of speech to be generated.
The pulse rate of 131Hz used in Table \ref{eff_table2} is generally
representative being the mean value observed in $F_0$ tracks extracted from the
VCTK training data. However, even under the worst pathological behaviour
(all frames voiced and spoken at 400Hz) complexity of system $P$ is still only
322.4 MFLOPS. A desire to limit such variability was part of the motivation to
have only a small number of layers operating at the pulse rate in Puffin.

We can see that the proposed vocoder -- despite operating at over twice the sampling rate -- requires substantially fewer operations per second
even when compared to HiFi-GAN version 3 whose smaller generator is intended
for wide deployment (188 vs.\ 3873). Similar comparisons hold with other widely-used systems prioritising efficiency, such as LPCNet \cite{valin2019lpcnet}. Eq.\ 8 in \cite{valin2019lpcnet} (modified to include frame-level terms) gives 2292 MFLOPS with the configuration used there, and 1332 with $N_A$ (width of first GRU layer) reduced from 384 to 256. Again, the system we propose is more efficient, despite operating at three times LPCNet's 16kHz sample rate.

The real time factor of an entire TTS system, incorporating vocoder $P$ as described here as well as acoustic model etc.\ on a single core of a common choice of hardware for AAC devices (Intel Atom x5-Z8350; RAM: 4GB) is 0.45 to produce speech at 48kHz.
A comparable acoustic model combined with LPCNet (even with $N_A$ reduced to 256) is not viable on the same machine, where it runs at 1.46 real time to produce 16kHz speech.

\begin{table}[t!]
\footnotesize
\centering
\caption{Complexity of systems evaluated. \label{eff_table2}}
\vspace{5pt}
$H^1$: HiFi-GAN v1 \\
\begin{tabular}{l r r r r r r}  \hline
 & $i$ & $o$ & $k$ & $l$ & rate (hz) & MFLOPS \\
 \hline
Conv & 80 & 512 & 7 & - & 86 & 49.3 \\
Upsa & 512 & 256 & 16 & - & 86 & 360.7 \\
Resi & 256 & 256 & 3 & 6 & 689 & 1625.6 \\
Resi & 256 & 256 & 7 & 6 & 689 & 3793.0 \\
Resi & 256 & 256 & 11 & 6 & 689 & 5960.4 \\
Upsa & 256 & 128 & 16 & - & 689 & 722.5 \\
Resi & 128 & 128 & 3 & 6 & 5512 & 3251.1 \\
Resi & 128 & 128 & 7 & 6 & 5512 & 7585.9 \\
Resi & 128 & 128 & 11 & 6 & 5512 & 11920.7 \\
Upsa & 128 & 64 & 4 & - & 5512 & 361.2 \\
Resi & 64 & 64 & 3 & 6 & 11025 & 1625.7 \\
Resi & 64 & 64 & 7 & 6 & 11025 & 3793.3 \\
Resi & 64 & 64 & 11 & 6 & 11025 & 5960.9 \\
Upsa & 64 & 32 & 4 & - & 11025 & 180.6 \\
Resi & 32 & 32 & 3 & 6 & 22050 & 812.9 \\
Resi & 32 & 32 & 7 & 6 & 22050 & 1896.7 \\
Resi & 32 & 32 & 11 & 6 & 22050 & 2980.5 \\
Conv & 32 & 1 & 7 & - & 22050 & 9.9 \\
 & &&&&&                    \textbf{52890.8} \\
\hline
\end{tabular}

\vspace{5pt}
$H^3$: HiFi-GAN v3 \\
\begin{tabular}{l r r r r r r}  \hline
 & $i$ & $o$ & $k$ & $l$ & rate (hz) & MFLOPS \\
 \hline
Conv & 80 & 256 & 7 & - & 86 & 24.7 \\
Upsa & 256 & 128 & 16 & - & 86 & 90.2 \\
Resi & 128 & 128 & 3 & 2 & 689 & 135.5 \\
Resi & 128 & 128 & 5 & 2 & 689 & 225.8 \\
Resi & 128 & 128 & 7 & 2 & 689 & 316.1 \\
Upsa & 128 & 64 & 16 & - & 689 & 180.6 \\
Resi & 64 & 64 & 3 & 2 & 5512 & 270.9 \\
Resi & 64 & 64 & 5 & 2 & 5512 & 451.5 \\
Resi & 64 & 64 & 7 & 2 & 5512 & 632.2 \\
Upsa & 64 & 32 & 8 & - & 5512 & 180.6 \\
Resi & 32 & 32 & 3 & 2 & 22050 & 271.0 \\
Resi & 32 & 32 & 5 & 2 & 22050 & 451.6 \\
Resi & 32 & 32 & 7 & 2 & 22050 & 632.2 \\
Conv & 32 & 1 & 7 & - & 22050 & 9.9 \\
 & &&&&&                    \textbf{3872.6} \\
\hline
\end{tabular}

\vspace{5pt}
$P$: Puffin standard \\
\begin{tabular}{l r r r r r r}  \hline
 & $i$ & $o$ & $k$ & $d$ & rate (hz) & MFLOPS \\
 \hline
Conv & 32 & 256 & 3 & 1.0 & 100 & 4.9 \\
Conv & 256 & 256 & 3 & 1.0 & 100 & 39.3 \\
Conv & 256 & 256 & 3 & 1.0 & 100 & 39.3 \\
Conv & 256 & 256 & 3 & 1.0 & 100 & 39.3 \\
Conv & 256 & 256 & 3 & 1.0 & 131 & 51.5 \\
Conv & 256 & 2064 & 1 & 0.1 & 131 & 13.8 \\
 & &&&&&                    \textbf{188.2} \\
\hline
\end{tabular}

\vspace{5pt}
$P^{L}$: Puffin large \\
\begin{tabular}{l r r r r r r}  \hline
 & $i$ & $o$ & $k$ & $d$ & rate (hz) & MFLOPS \\
 \hline
Conv & 32 & 1024 & 3 & 1.0 & 100 & 19.7 \\
Conv & 1024 & 1024 & 3 & 1.0 & 100 & 629.1 \\
Conv & 1024 & 1024 & 3 & 1.0 & 100 & 629.1 \\
Conv & 1024 & 1024 & 3 & 1.0 & 100 & 629.1 \\
Conv & 1024 & 1024 & 3 & 1.0 & 131 & 824.2 \\
Conv & 1024 & 2064 & 1 & 1.0 & 131 & 553.7 \\
 & &&&&&                    \textbf{3285.0} \\
\hline
\end{tabular}
\vspace{-22pt}
\end{table}

\vspace{-10pt}
\subsection{Listening experiment design}
\vspace{-5pt}
We performed a MUSHRA-style listening test where participants were asked to rate
the quality of the generated samples. In total we generated 9 sentences per
speaker in the test set (in total 54 sentences); all sentences contained
different text. Sentences were randomly selected from the set each speaker
recorded. The sentences were then divided equally across 3 different listening
test, such that each test contained 18 number of sentences per speaker. Each
participant was randomly assigned to a specific test; i.e. each participant was
presented with 18 number of MUSHRA screens/sentences. The test duration was
around 20mins. 

Participants were recruited via Prolific Academic using the following screening
criteria: between 18 and 50 years old, no known hearing impairement or
difficulties, and English as a first language. Participants were asked to
perform the test in a quiet place wearing headphones. 20 participants took part
in the study.

\vspace{-10pt}
\subsection{Results}
\vspace{-5pt}

\begin{figure}
\includegraphics[clip, trim=1.5cm 7.5cm 1cm 8cm, width=0.5\textwidth]{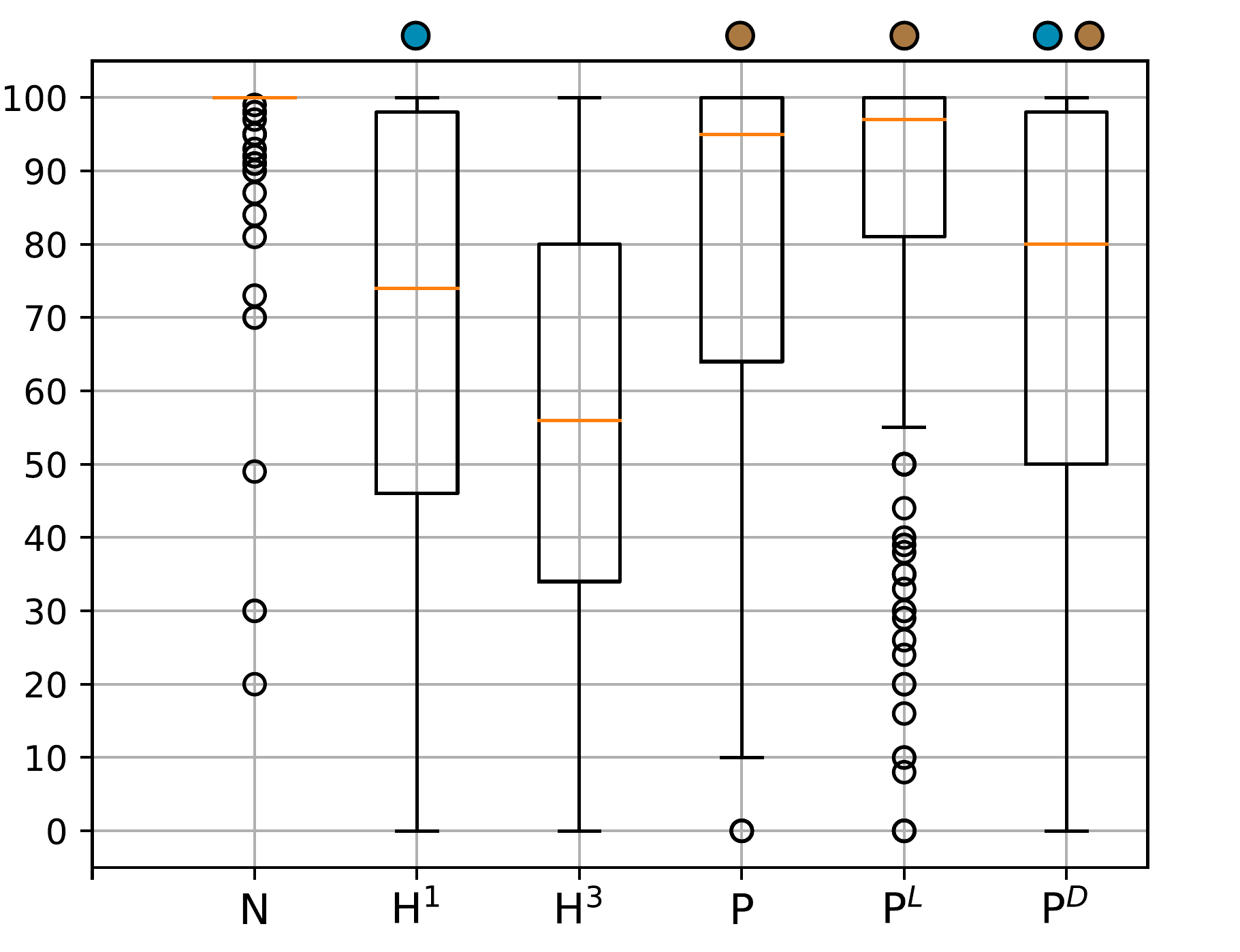}
\vspace{-22pt}
\caption{Boxplot of listening test scores.} \label{fig:boxplot}
\vspace{-20pt}
\end{figure}

Figure\ \ref{fig:boxplot} shows the results of the listening test in the form of a
boxplot. Orange lines indicate the mean rating for each system. To compute
significance, a Wilcoxon signed-rank test was used (significance level = 0.05),
with a Bonferroni correction for multiple pairwise comparisons. Following the
standard exclusion criteria in \cite{Mushra}, participants were discarded if
they rated the hidden natural reference less than 90 more than 15\% of times,
which resulted in 14 valid participants (out of 20). A significant difference
between the ratings for H$^3$ and all other systems was found, as well as for N
and all other systems; H$^1$ was significantly different from all systems
except P$^D$ (indicated by blue dots). For the various proposed systems, P was
not rated significantly different from the rest (indicated by yellow dots),
while P$^D$ and P$^L$ were significantly different from each other.

The higher rating of system P (the proposed system at 48kHz) than that of system
P$^D$ (the proposed system downsampled to 22kHz) justifies the generation of
fullband speech. System H$^1$
(the strong benchmark) is not significantly different from the downsampled
proposed system. We suggest that this inconsistency in ratings of speech at
difference sample rates requires more evaluation, potentially looking at
whether people consistently rate natural speech at these two sampling rates as
different.

The ratings of P and P$^L$ are similar. This indicates that the extra complexity
of P$^L$ is not required, although some benefit is suggested.

Finally, H$^3$ is rated as significantly worse than all other systems. Informal
listening of samples generated by this system suggests that this model suffers
from some periodicity/voicing artefacts, which are not present in the other
systems.

\section{Conclusions}
\vspace{-6pt}

In this work we have presented a new waveform generation method combining
traditional signal processing methods from an older generation of vocoders with new deep learning approaches. Our results, firstly,
suggest that it is worth pursuing fullband speech, despite the current focus
on e.g. 22kHz audio in neural vocoder research. More importantly, we have found
that our system is able to generate speech whose quality matches that of strong
baseline systems, at a fraction of the run time computational cost. This makes
the deployment of high quality TTS voices on low-powered devices (such that are
often used by people with AAC needs) feasible, meaning a population of users
whose communication might be solely carried out using TTS could also benefit
from recent quality improvements.

\label{sec:ref}
\bibliographystyle{IEEEbib}
\bibliography{references}

\begin{thebibliography}{10}

\bibitem{Masanori16}
Masanori Morise, Fumiya Yokomori, and Kenji Ozawa,
\newblock ``{WORLD: A Vocoder-Based High-Quality Speech Synthesis System for
  Real-Time Applications},''
\newblock {\em IEICE Trans. Inf. Syst.}, vol. E99.D, no. 7, 2016.

\bibitem{Espic17}
Felipe {Espic}, Cassia {Valentini-Botinhao}, and Simon King,
\newblock ``Direct modelling of magnitude and phase spectra for statistical
  parametric speech synthesis,''
\newblock in {\em Proc. Interspeech}, 2017, pp. 1383--1387.

\bibitem{mehri2016samplernn}
Soroush Mehri, Kundan Kumar, Ishaan Gulrajani, Rithesh Kumar, Shubham Jain,
  Jose Sotelo, Aaron Courville, and Yoshua Bengio,
\newblock ``Sample{RNN}: An unconditional end-to-end neural audio generation
  model,''
\newblock in {\em Proc. ICLR}, 2017.

\bibitem{shen2017natural}
Jonathan Shen, Ruoming Pang, Ron~J. Weiss, Mike Schuster, Navdeep Jaitly,
  Zongheng Yang, Zhifeng Chen, Yu~Zhang, Yuxuan Wang, RJ~Skerry Ryan, Rif~A.
  Saurous, Yannis Agiomyrgiannakis, and Yonghui Wu,
\newblock ``Natural {TTS} synthesis by conditioning {W}avenet on {MEL}
  spectrogram predictions,''
\newblock in {\em {Proc. ICASSP}}, 2018, pp. 4779 -- 4783.

\bibitem{matsubara2022comparison}
Keisuke Matsubara, Takuma Okamoto, Ryoichi Takashima, Tetsuya Takiguchi, Tomoki
  Toda, and Hisashi Kawai,
\newblock ``{Comparison of real-time multi-speaker neural vocoders on CPUs},''
\newblock {\em Acoustical Science and Technology}, vol. 43, no. 2, pp.
  121--124, 2022.

\bibitem{oord2016wavenet}
Aäron van~den Oord, Sander Dieleman, Heiga Zen, Karen Simonyan, Oriol Vinyals,
  Alexander Graves, Nal Kalchbrenner, Andrew Senior, and Koray Kavukcuoglu,
\newblock ``Wave{N}et: A generative model for raw audio,''
  https://arxiv.org/abs/1609.03499, 2016,
\newblock arXiv:1609.03499.

\bibitem{oord2018parallel}
Aäron van~den Oord, Yazhe Li, Igor Babuschkin, Karen Simonyan, Oriol Vinyals,
  Koray Kavukcuoglu, George van~den Driessche, Edward Lockhart, Luis Cobo,
  Florian Stimberg, Norman Casagrande, Dominik Grewe, Seb Noury, Sander
  Dieleman, Erich Elsen, Nal Kalchbrenner, Heiga Zen, Alex Graves, Helen King,
  Tom Walters, Dan Belov, and Demis Hassabis,
\newblock ``Parallel {W}ave{N}et: Fast high-fidelity speech synthesis,''
\newblock in {\em Proc. ICML}, Jul 2018, pp. 3918--3926.

\bibitem{prenger2019waveglow}
Ryan Prenger, Rafael Valle, and Bryan Catanzaro,
\newblock ``Waveglow: A flow-based generative network for speech synthesis,''
\newblock in {\em Proc. ICASSP}, 2019, pp. 3617--3621.

\bibitem{kumar2019melgan}
Kundan Kumar, Rithesh Kumar, Thibault de~Boissiere, Lucas Gestin, Wei~Zhen
  Teoh, Jose Sotelo, Alexandre de~Brebisson, Yoshua Bengio, and Aaron
  Courville,
\newblock ``{MelGAN}: Generative adversarial networks for conditional waveform
  synthesis,'' http://arxiv.org/abs/1910.06711, 2019,
\newblock arXiv:1910.06711.

\bibitem{hifigan}
Jungil Kong, Jaehyeon Kim, and Jaekyoung Bae,
\newblock ``{HiFi-GAN}: Generative adversarial networks for efficient and high
  fidelity speech synthesis,''
\newblock in {\em Proc. {NeurIPS}}, 2020, vol.~33, pp. 17022--17033.

\bibitem{valin2019lpcnet}
Jean{-}Marc Valin and Jan Skoglund,
\newblock ``{LPCNET:} improving neural speech synthesis through linear
  prediction,''
\newblock in {\em Proc. ICASSP}, 2019, pp. 5891--5895.

\bibitem{cui2020efficient}
Yang Cui, Xi~Wang, Lei He, and Frank~K. Soong,
\newblock ``An efficient subband linear prediction for {LPCN}et-based neural
  synthesis,''
\newblock in {\em Proc. Interspeech}, 2020, pp. 3555--3559.

\bibitem{kaneko2022istftnet}
Takuhiro Kaneko, Kou Tanaka, Hirokazu Kameoka, and Shogo Seki,
\newblock ``{ISTFTNET: Fast and Lightweight Mel-Spectrogram Vocoder
  Incorporating Inverse Short-Time Fourier Transform},''
\newblock in {\em Proc. ICASSP}, 2022, pp. 6207--6211.

\bibitem{webber2023autovocoder}
Jacob~J Webber, Cassia Valentini-Botinhao, Evelyn Williams, Gustav~Eje Henter,
  and Simon King,
\newblock ``Autovocoder: Fast waveform generation from a learned speech
  representation using differentiable digital signal processing,''
\newblock in {\em Proc. ICASSP}, 2023.

\bibitem{wangNSF}
Xin Wang, Shinji Takaki, and Junichi Yamagishi,
\newblock ``Neural source-filter waveform models for statistical parametric
  speech synthesis,''
\newblock {\em IEEE/ACM Trans. Audio, Speech and Lang. Proc.}, vol. 28, pp.
  402--415, 2020.

\bibitem{yamamoto2020parallel}
Ryuichi Yamamoto, Eunwoo Song, and Jae-Min Kim,
\newblock ``Parallel wavegan: A fast waveform generation model based on
  generative adversarial networks with multi-resolution spectrogram,''
\newblock in {\em Proc. ICASSP}, 2020, pp. 6199--6203.

\bibitem{vctk}
Junichi Yamagishi, Christophe Veaux, and Kirsten MacDonald,
\newblock ``{CSTR VCTK} corpus: English multi-speaker corpus for {CSTR} voice
  cloning toolkit,'' \url{https://doi.org/10.7488/ds/1994}, 2017.

\bibitem{NickData}
Cassia Valentini-Botinhao, Catherine Mayo, and Martin Cooke,
\newblock ``Hurricane natural speech corpus - higher quality version,''
  \url{https://doi.org/10.7488/ds/2482}, 2019.

\bibitem{AlbaData}
Cassia Valentini-Botinhao and Junichi Yamagishi,
\newblock ``Alba speech corpus,'' \url{https://doi.org/10.7488/ds/2506}, 2019.

\bibitem{sv56}
International Telecommunication Union Radiocommunication Assembly,
\newblock {\em International Telecommunication Union, Recommendation G.191:
  Software Tools and Audio Coding Standardization,}, November 2005.

\bibitem{Reaper}
``{REAPER: Robust Epoch And Pitch EstimatoR},''
  \url{https://github.com/google/REAPER}, 2017.

\bibitem{Mushra}
International Telecommunication Union Radiocommunication Assembly, Geneva,
  Switzerland,
\newblock {\em Method for the subjective assessment of intermediate quality
  level of coding systems}, March 2003.

\end{thebibliography}

\end{document}